\documentclass[conference]{IEEEtran}
\usepackage[pdftex]{graphicx}
\usepackage{amsfonts}
\usepackage{amsmath}
\usepackage{multicol}
\usepackage{adjustbox}
\usepackage{cite}
\usepackage{footnote}
\usepackage[bottom]{footmisc}
\usepackage{lipsum}
\newcommand\blfootnote[1]{%
	\begingroup
	\renewcommand\thefootnote{}\footnote{#1}%
	\addtocounter{footnote}{-1}%
	\endgroup
}
\usepackage{subdepth}
\usepackage{bm}
\usepackage{setspace}
\usepackage{scalerel} %for low font size of subscript
\DeclareGraphicsExtensions{.pdf}
\newcommand{\be}{\begin{equation}}
\newcommand{\eq}{\end{equation}}
\DeclareMathOperator{\EX}{\mathbb{E}}% expected value

\IEEEoverridecommandlockouts

\usepackage{stfloats}
\usepackage{graphicx}
\usepackage{epstopdf}
\usepackage{caption}
\usepackage{subcaption}
\usepackage{amsmath}
\usepackage{amssymb}
\usepackage{float}
\usepackage{adjustbox}
\usepackage{booktabs}
\usepackage{array}
\usepackage{makecell}
\usepackage{placeins}
\usepackage{siunitx}
\usepackage{mathtools}
\usepackage{lettrine}
\usepackage{balance}
\usepackage{lipsum}
\usepackage{xcolor}
\usepackage{flushend}
\DeclareMathAlphabet\mathbfcal{OMS}{cmsy}{b}{n} % bold mathcal font
\DeclareMathOperator*{\argmax}{arg\,max}
\usepackage[linesnumbered,ruled,vlined]{algorithm2e} %for algorithms
\SetAlFnt{\small}
\makeatletter
\newcommand{\nosemic}{\renewcommand{\@endalgocfline}{\relax}}% Drop semi-colon ;
\newcommand{\dosemic}{\renewcommand{\@endalgocfline}{\algocf@endline}}% Reinstate semi-colon ;
\let\oldnl\nl% Store \nl in \oldnl
\newcommand{\nonl}{\renewcommand{\nl}{\let\nl\oldnl}}% Remove line number for one line
\makeatother

\usepackage[font=scriptsize]{caption}

\usepackage{geometry}
\newcommand*\squeezespaces[1]{% %% <- #1 is a number between 0 and 1
	\thickmuskip=\scalemuskip{\thickmuskip}{#1}%
	\medmuskip=\scalemuskip{\medmuskip}{#1}%
	\thinmuskip=\scalemuskip{\thinmuskip}{#1}%
	\nulldelimiterspace=#1\nulldelimiterspace
	\scriptspace=#1\scriptspace
} 
\newcommand*\scalemuskip[2]{%
	\muexpr #1*\numexpr\dimexpr#2pt\relax\relax/65536\relax
} %% <- based on  https://tex.stackexchange.com/a/198966/156366

\def\BibTeX{{\rm B\kern-.05em{\sc i\kern-.025em b}\kern-.08em
		T\kern-.1667em\lower.7ex\hbox{E}\kern-.125emX}}
\newcommand{\rmnum}[1]{\uppercase\expandafter{\romannumeral #1\relax}}
\begin{document}
\bstctlcite{IEEEexample:BSTcontrol}

\title{Multiple UAV-Assisted Cooperative DF Relaying in Multi-User Massive MIMO IoT Systems \vspace{-1.5ex}}
\author{Mobeen Mahmood, Yicheng Yuan, Tho Le-Ngoc \\ Department of Electrical and Computer Engineering, McGill University, Montreal, QC, Canada \\
	Email: mobeen.mahmood@mail.mcgill.ca, yicheng.yuan@mail.mcgill.ca, tho.le-ngoc@mcgill.ca 
	\vspace{-4.5ex} 
}

\date{}
\maketitle
\begin{abstract}
This work considers a multi-user massive multiple-input multiple-output (MU-mMIMO) Internet-of-Things (IoT) system, where multiple unmanned aerial vehicles (UAVs) operating as decode-and-forward (DF) relays connect the base station (BS) to a large number of IoT devices. To maximize the total achievable rate, we propose a novel joint optimization problem of hybrid beamforming (HBF), multiple UAV relay positioning, and power allocation (PA) to multiple IoT users. The study adopts a geometry-based millimeter-wave (mmWave) channel model for both links and utilizes sequential optimization based on K-means UAV-user association. The radio frequency (RF) stages are designed based on the slow time-varying angular information, while the baseband (BB) stages are designed utilizing the reduced-dimension effective channel matrices. The illustrative results show that multiple UAV-assisted cooperative relaying systems outperform a single UAV system in practical user distributions. Moreover, compared to fixed positions and equal PA of UAVs and BS, the joint optimization of UAV location and PA substantially enhances the total achievable rate. 
\vspace{-2.5ex}
\end{abstract}

\blfootnote{This work was supported in part by Huawei Technologies Canada and in part by the Natural Sciences and Engineering Research Council of Canada.}
%\begin{IEEEkeywords}
%	Unmanned aerial vehicles, massive multiple-input multiple-output, hybrid beamforming, dual-hop communications, decode-and-forward relay.
%\end{IEEEkeywords}
\vspace{-1.2ex}
\section{Introduction}
\vspace{-1ex}
\IEEEPARstart{R}{elaying} has long been considered as a potent means to mitigate signal attenuation and fading effects, thereby extending the coverage and reliability of wireless systems.
% In the context of massive multiple-input multiple-output (mMIMO) Internet-of-Things (IoT), where base stations (BSs) are endowed with a multitude of antennas, relaying emerges as a strategic enabler to further optimize spatial diversity and multiplexing gains, thereby elevating the performance of wireless networks \cite{uyoata2021relaying}. 
%As we look ahead to the development of 6G wireless networks, the role of Internet-of-Things (IoT) becomes increasingly prominent. IoT is expected to play a pivotal role in shaping these future networks, with projections indicating the integration of tens of billions of IoT devices into wireless networks. These devices will facilitate various applications, such as industrial communications, e-healthcare, and smart city infrastructure. However, the broad scope and growing influence of IoT applications introduce new challenges of coverage and capacity requirements in wireless communication systems \cite{lv2018optimization,uyoata2021relaying,lyu2018relay}. \par 
The integration of \textit{relaying} techniques in multi-user massive multiple-input multiple-output (MU-mMIMO) Internet-of-Things (IoT) systems is a practical consideration to further optimize spatial diversity and multiplexing gains, thereby elevating the performance of wireless networks \cite{uyoata2021relaying}. In this regard, unmanned aerial vehicles (UAVs) as a moving relay can offer the following advantages: 1) high mobility, which allows for the dynamic adjustment of their locations to optimize communication conditions \cite{delay_UAV}; 2) ease of deployment and low energy consumption; and 3) capability to reach inaccessible locations. Thus, UAV-aided communications is a viable option for future MU-mMIMO IoT applications, addressing coverage issues and minimizing communications system overhead \cite{feng2018uav}.\par 
The high altitude deployment of UAVs increases the likelihood of line-of-sight (LoS) dominated air-to-ground communication channels. While placing UAVs at optimal locations is essential to improve channel quality, efficient interference mitigation schemes are also required to enhance the network capacity. The three-dimensional (3D) beamforming capability of MU-mMIMO can effectively suppress interference among IoT nodes \cite{UAV_Yi2021}. In this regard, two common approaches for alleviating interference are fully-digital beamforming (FDBF) and hybrid beamforming (HBF). FDBF requires radio frequency (RF) chains equal to the number of antenna elements, which hinders its implementation in UAV-assisted IoT systems due to the prohibitive cost, complexity, and limited power supply of UAVs. Conversely, HBF involves the design of both the RF-stage and baseband (BB)-stage, and can approach the performance of FDBF by reducing the number of power-hungry RF chains, thereby improving energy efficiency \cite{mahmood2021energy, mobeen_3D,mobeen2020}. Recent research studies have focused on optimizing UAV location with particular emphasis on HBF solutions to maximize throughput or minimize transmit power \cite{Mobeen_UAV1,uav_du2020,Mobeen_UAV_Spherical,UAV_relay_IoT_1}. In particular, \cite{Mobeen_UAV1,uav_du2020} investigate the joint optimization of UAV deployment, while considering HBF at BS and UAV for maximum sum-rate. An amply-and-forward UAV relay with analog beamforming architecture is considered in \cite{Mobeen_UAV_Spherical} to maximize the capacity in a dual-hop mMIMO IoT system. Similarly, \cite{UAV_relay_IoT_1} optimizes the source/relay power allocation and UAV trajectory to enhance end-to-end system throughput.\par 
Most existing studies consider a single UAV (for instance \cite{Mobeen_UAV1,uav_du2020,Mobeen_UAV_Spherical,UAV_relay_IoT_1}), which can provide only limited user coverage and access. On the other hand, a network of multiple UAVs can efficiently enlarge the coverage region and increase the number of served users\cite{chen2018multiple,UAV_Chen2021,nouri2023}. The authors in \cite{chen2018multiple} propose a multi-UAV relaying system, and compared the performance of a single multi-hop link and multiple dual-hop links. In \cite{UAV_Chen2021}, the position of multiple UAV-mounted BS is optimized to enhance the coverage area, while satisfying quality-of-service (QoS) requirements. Similarly, the placement and the number of UAVs are optimized in \cite{nouri2023}, while adhering to network capacity and coverage constraints. Research works such as \cite{Mobeen_UAV1,uav_du2020,Mobeen_UAV_Spherical,UAV_relay_IoT_1,chen2018multiple,UAV_Chen2021,nouri2023} have ignored the direct link between the BS and IoT users. However, it is well known that when direct link is non-negligible or not too weak, the spatial diversity can be enhanced via direct and cooperative multiple path (through UAV relays). In practice, assuming no direct link simplifies the design of the joint source-relay beamforming. When the direct link is involved, the received signals via source-to-destination link and multiple UAV relay-to-destination links are combined at the destination (IoT user) to enhance the overall signal strength. As such, the beamforming vector design at the BS needs to take into account both of these links. Furthermore, the optimization problem becomes more complex. \par 
To solve this challenging issue, we consider a more practical cooperative transmission approach, integrating both the direct link from BS to IoT devices and the indirect links via multiple UAV relays to IoT users in MU-mMIMO IoT systems to enhance the capacity and overcome the coverage issues. Our objectives here are twofold: first, to show that the use of multiple UAVs as relays can significantly increase the sum-rate capacity; and second, the optimization of UAV location, user association, and power allocation (PA) at BS and UAVs combined with HBF design can provide better performance than fixed UAV locations with equal PA. To the best of our knowledge, the joint optimization of multiple UAV placements, PA at the BS, and the design of HBF for both the BS and UAVs is an unaddressed problem, presenting a significant opportunity to advance the field of UAV-assisted MU-mMIMO IoT communications networks. The joint optimization problem is highly non-convex, therefore, we utilize structured sequential optimization to address the multi-faceted optimization problem by splitting it into two subproblems. First, K-means-based user clustering is used for UAV-users association based on the 3D geometry-based millimeter-wave (mmWave) channel model. Then, the location of each UAV is optimized jointly with PA using swarm intelligence. The RF beamforming stages for BS and UAVs are designed based on the slow time-varying angle-of-departure (AoD)/angle-of-arrival (AoA) information, and BB stages are formulated using the reduced-dimensional effective channel matrices. The illustrative results show that the proposed joint optimization scheme for multiple UAV-assisted relaying significantly enhances the performance in MU-mMIMO IoT systems.
%--------- Section II ---------------------
\vspace{-1ex}
\section{System and Channel Model}
\vspace{-1ex}
In this section, we introduce the system and channel models of the proposed multi-UAV-assisted relaying and the HBF design for a multiple dual-hop MU-mMIMO IoT system.
\vspace{-1ex}
\subsection{System Model}
\begin{figure}[!t] 
	\centering
	\subfloat[\label{fig:fig1a}]{% 
		\includegraphics[height=4.2cm,width=\columnwidth]{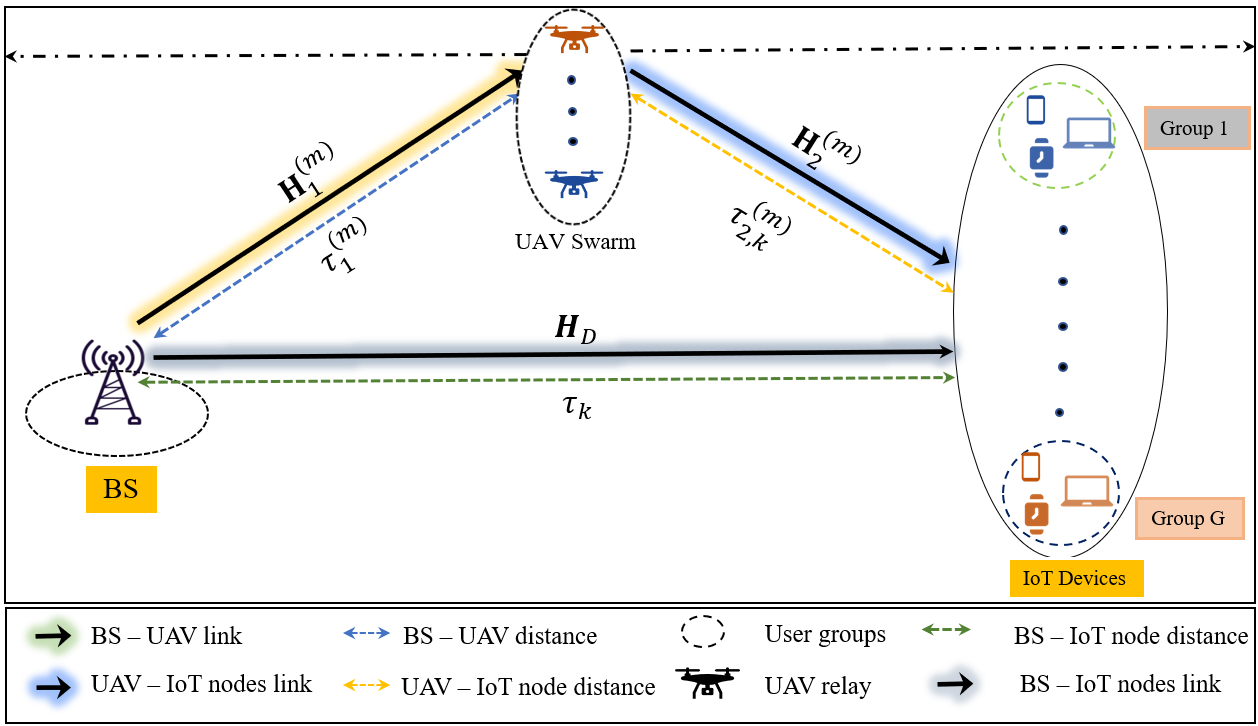} 
	} \hfil 
	\subfloat[\label{fig:fig1b}]{% 
		\includegraphics[height=3.5cm,width=\columnwidth]{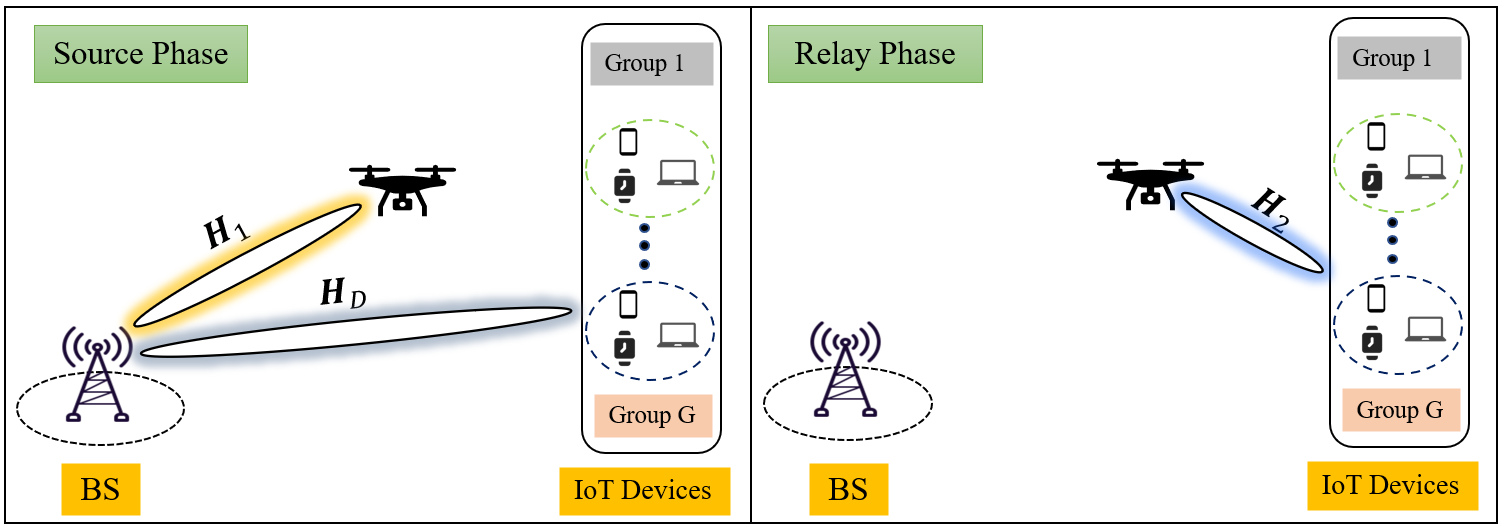} 
	} 
	\caption{Multiple UAV-assisted MU-mMIMO IoT communications. (a) network model. (b) UAV as DF relay transmission phases.}
	\label{fig:fig1}
	\vspace{-1ex}
\end{figure}

\vspace{-0.5ex}
We consider a downlink MU-mMIMO IoT network, where a large number of non-overlapping IoT devices are connected to an IoT gateway via wire or wireless links. Due to severe shadowing and blocking effect, many ground IoT devices experience low signal-quality from the BS/eNodeB that is equipped with a large array having $N_b$ antenna elements. To address this challenging scenario, we consider a cooperative relaying system model as shown in Fig. \ref{fig:fig1}(a), to serve $\mathbb{K} = \{1, \cdots, K\}$ single-antenna IoT nodes, which are clustered in $G$ groups, where $g^{th}$ group has $K_g$ IoT nodes such that $K = \sum_{g=1}^{G}K_g$. Then, $M$ different UAVs, indexed by the set $\mathbb{U} = \{u_1, u_2, \cdots, u_M\}$, are deployed to serve $K = M K_m$ IoT users, where $K_m$ is number of users served by $m^{th}$ UAV, which operates as DF relay between BS/eNodeB and IoT node\footnote{We assume equal UAV-user clustering for simplicity. However, it can be applicable for unequal user clustering, which is left as our future work.}. Let $\left(x_b, y_b, z_b\right)$, $(x_u^{(m)}, y_u^{(m)}, z_u^{(m)})$ and $\left(x_k, y_k, z_k \right)$ denote the locations of BS, $m^{th}$ UAV relay, and $k^{th}$ IoT user, respectively. Then, we define the 3D distances for multiple UAV-assisted MU-mMIMO IoT system as follows: $\tau_1^{(m)}$ = $\sqrt{(x_u^{(m)} - x_b)^2 + (y_u^{(m)} - y_b)^2 + (z_u^{(m)} - z_b)^2}$, $\tau_{2,k}^{(m)}$ = $\sqrt{(x_u^{(m)} - x_k)^2 + (y_u^{(m)} - y_k)^2 + (z_u^{(m)} - z_k)^2}$, $\tau_{k}$ = $\sqrt{(x_b - x_k)^2 + (y_b - y_k)^2 + (z_b - z_k)^2}$, where $\tau_1^{(m)}$, $\tau_{2_k}^{(m)}$ and $\tau_{k}$ are the 3D distance between $m^{th}$ UAV \& BS, between $m^{th}$ UAV and $k^{th}$ IoT node, and between BS and $k^{th}$ IoT node, respectively. Each UAV is equipped with $N_{r} (N_{t})$ antennas for receiving (transmitting) signals from (to) BS (IoT users). For simplicity, we assume a homogeneous fleet of UAVs with consistent specifications and functionality. Unlike traditional static relaying, which uses fixed relay locations, we presumptively use multiple UAVs as a movable relays. \par
\begin{figure*}[!t]
%	\captionsetup{justification=centering}
	\includegraphics[height= 4cm, width=2\columnwidth]{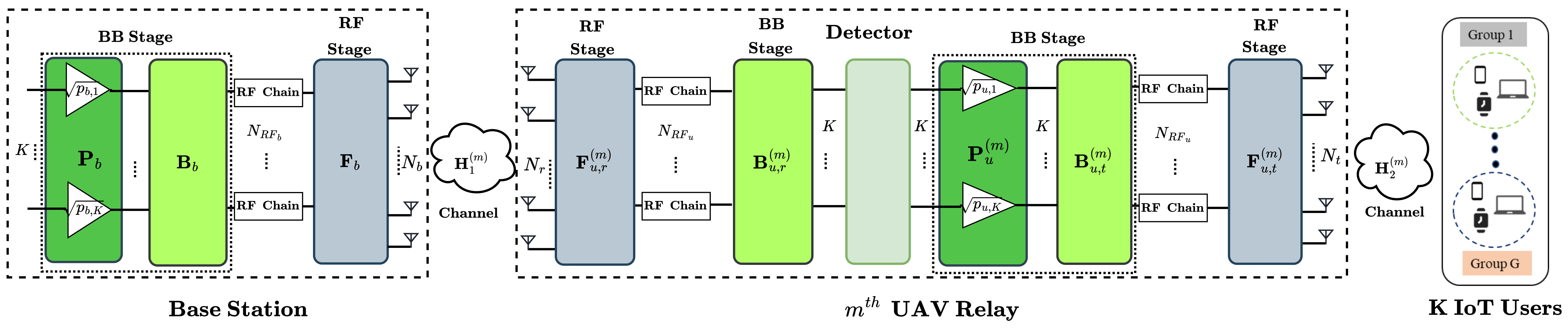} 
	\caption{Multiple UAV-assisted MU-mMIMO HBF system model.}
	\label{fig:fig2}
	\vspace{-4ex}
\end{figure*} 
For the downlink transmission of $N_S = K$ data streams, we consider HBF for BS and all UAV relays as shown in Fig. \ref{fig:fig2}. The BS consists of RF beamforming stage $\mathbf{F}_b$ $\in$ $\mathbb{C}^{N_b \times N_{{RF}_b}}$, BB stage $\mathbf{B}_b$ $\in$ $\mathbb{C}^{N_{{RF}_b} \times K}$, and MU PA matrix $\mathbf{P}_b = \text{diag}(\sqrt{p_{b_1}}, \cdots, \sqrt{p_{b_K}})$ $\in$ $\mathbb{C}^{K \times K}$. Here, $N_{{RF}_b}$ is the RF chains such that $N_S \leq N_{{RF}_b} \leq N_b$ to guarantee multi-stream transmission. We consider half-duplex (HD) DF relaying for each UAV. Therefore,
each round of information transmission from BS to IoT nodes can be divided into two phases: 1) source phase (SP); and 2) relay phase (RP) as illustrated in Fig. \ref{fig:fig1}(b).
%\footnote{The use of full-duplex (FD) relays in multiple UAV-assisted mMIMO IoT system is beyond the scope of this paper and is left as our future work.}.
In SP, BS transmits $K$ data streams to the following: 1) $K$ IoT users through channel $\mathbf{H}_D$ $\in$ $\mathbb{C}^{K \times N_b}$; and 2) each $m^{th}$ UAV via channel $\mathbf{H}_1^{(m)}$ $\in$ $\mathbb{C}^{N_{r} \times N_b}$. Using $N_{r}$ antennas, each UAV receives signals with RF stage $\mathbf{F}_{u,r}^{(m)}$ $\in$ $\mathbb{C}^{N_{{RF}_{u}} \times N_{r}}$ and BB stage $\mathbf{B}_{u,r}^{(m)}$ $\in$ $\mathbb{C}^{K \times N_{{RF}_{u}}}$. In RP, each UAV decodes the received information and then forwards the decoded information to $K$ IoT nodes using RF beamformer $\mathbf{F}_{u,t}^{(m)} \in$ $\mathbb{C}^{N_{t} \times N_{{RF}_{u}}}$, BB stage $\mathbf{B}_{u,t}^{(m)} \in$ $\mathbb{C}^{N_{{RF}_{u}} \times K}$, and MU PA matrix $\mathbf{P}_u^{(m)} = \text{diag}(\sqrt{p_{u_1}^{(m)}}, \cdots, \sqrt{p_{u_K}^{(m)}})$ $\in$ $\mathbb{C}^{K \times K}$ through channel $\mathbf{H}_2^{(m)}$ $\in$ $\mathbb{C}^{K \times N_{t}}$, where $p_{u_k}^{(m)}$ reflects the allocated power to $k^{th}$ user from $m^{th}$ UAV. The implementation of all RF beamforming/combining stages involves the use of phase-shifters (PSs) and thus, impose a constant-modulus (CM) constraint, i.e., $|\mathbf{F}_b (i,j)| = \frac{1}{\sqrt{N_b}}$, $|\mathbf{F}_{u,r}^{(m)} (i,j)| = \frac{1}{\sqrt{N_{r}}}$, $|\mathbf{F}_{u,t}^{(m)} (i,j)| = \frac{1}{\sqrt{N_{t}}} \hspace{1ex} \forall i,j,m$. For the data signal $\mathbf{d} = [d_1, d_2, \cdots, d_{K}]^T$ with $\EX \{\mathbf{d}\mathbf{d}^H\} = \mathbf{I}_{K}$ $\in$ $\mathbb{C}^{K \times K}$, the signal received at $m^{th}$ UAV (after BB processing) during SP is given as follows: \vspace{-0.5ex}
\begin{equation}
	\tilde{\mathbf{y}}_u^{(m)} = \mathbf{B}_{u,r}^{(m)}\mathbf{F}_{u,r}^{(m)}\mathbf{H}_1^{(m)} \mathbf{F}_b\mathbf{B}_b \mathbf{P}_b \mathbf{d} + \mathbf{B}_{u,r}^{(m)}\mathbf{F}_{u,r}^{(m)}\mathbf{n}_u^{(m)}, \label{eq:1} \vspace{-0.5ex}
\end{equation}
where $n_u^{(m)}$ $\in$ $\mathbb{C}^{N_{u,r}}$ denotes the zero-mean complex circularly symmetric Gaussian noise vector at $m^{th}$ UAV relay with covariance matrix $\EX \{\mathbf{n}_u\mathbf{n}_u^H\} = \sigma_{n_u}^2 \mathbf{I}_{N_{u,r}}$ \hspace{-2ex} $\in$ $\mathbb{C}^{N_{u,r} \times N_{u,r}}$. Then, the signal transmitted by $m^{th}$ UAV during RP is given as:  \vspace{-0.5ex}
\begin{equation}
	\hat{\mathbf{s}}^{(m)} = \mathbf{F}_{u,t}^{(m)}\mathbf{B}_{u,t}^{(m)}\mathbf{P}_u^{(m)} \hat{\mathbf{d}}^{(m)}, \label{eq:2} \vspace{-1ex}
\end{equation} 
where $\mathbf{\hat{d}}^{(m)}$ is the re-encoded signal at $m^{th}$ UAV relay. Each IoT node receives signal from BS and each $m^{th}$ UAV during SP and RP, respectively. Then, the received signal at $k^{th}$ IoT node from BS and $m^{th}$ UAV can be written as:
\begin{equation}\label{eq:received_User}
	\begin{split}
		y_{k}^{(m)\hspace{-0.5ex}} \hspace{-0.6ex}&=\hspace{-0.6ex} y_{k,SP} + y_{k,RP}^{(m)}, \\
		&= \mathbf{h}_{D,k\hspace{-0.1ex}}^{T} \hspace{-0.25ex} \mathbf{F}_{b}\hspace{-0.1ex}\mathbf{B}_{b}\hspace{-0.15ex}\mathbf{P}_b\hspace{-0.15ex} \hat{d}_{k\hspace{-0.15ex}}\hspace{-0.2ex} + \hspace{-0.2ex}n_{D_{k}\hspace{-0.25ex}} \hspace{-1ex} + \hspace{-0.5ex} \mathbf{h}_{2,k}^{T(m)}\hspace{-0.15ex} \mathbf{F}_{u,t}^{(m)}\hspace{-0.15ex}\mathbf{B}_{u,t}^{(m)}\hspace{-0.15ex}\mathbf{P}_u^{(m)\hspace{-0.25ex}}\hspace{-0.15ex} \hat{d}_{k}^{(m)\hspace{-0.5ex}} \hspace{-0.5ex}+\hspace{-0.5ex} n_{2_{k}}^{(m)}\hspace{-1ex},  \\ \vspace{-4ex}
		&=\hspace{-0.5ex} \mbox{$\squeezespaces{0.5}\underbrace{\sqrt{p_{b_{k}}\hspace{-0.4ex}}\mathbf{h}_{D,k\hspace{-0.1ex}}^{T} \hspace{-0.25ex}\mathbf{F}_{b}\hspace{-0.15ex}\mathbf{B}_{b}\hspace{-0.15ex}\mathbf{P}_b\hspace{-0.15ex}\hat{d}_{k\hspace{-0.25ex}} \hspace{-0.65ex}+\hspace{-0.75ex} \sqrt{p_{u_{k}^{\hspace{-0.35ex}(m)}}\hspace{-0.4ex}}\mathbf{h}_{2,k}^{T(m)}\hspace{-0.15ex} \mathbf{F}_{u,t}^{(m)}\hspace{-0.15ex}\mathbf{B}_{u,t}^{(m)}\hspace{-0.15ex}\mathbf{P}_u^{(m)\hspace{-0.25ex}}\hspace{-0.15ex} \hat{d}_{k}^{(m)\hspace{-0.5ex}}}_{\text{Desired Signal}} + $}\\ &\mbox{$\squeezespaces{0.5}\underbrace{\hspace{-0.85ex}\sum_{\hat{k}=1}^{K}\hspace{-0.75ex}\sqrt{p_{b_{{\hat{k}}}}\hspace{-0.4ex}}\mathbf{h}_{D,k\hspace{-0.15ex}}^{T} \hspace{-0.25ex}\mathbf{F}_{b}\hspace{-0.15ex}\mathbf{B}_{b}\hspace{-0.15ex}\mathbf{P}_b\hspace{-0.15ex}\hat{d}_{{\hat{k}}\hspace{-0.25ex}} \hspace{-0.5ex} + \hspace{-0.5ex}\sum_{\hat{k}=1}^{K}\hspace{-0.6ex}\sqrt{\hspace{-0.35ex}p_{u_{k}^{(m)}}\hspace{-0.2ex}}\mathbf{h}_{2,k}^{T(m)}\hspace{-0.15ex} \mathbf{F}_{u,t}^{(m)}\hspace{-0.15ex}\mathbf{B}_{u,t}^{(m)}\hspace{-0.15ex}\mathbf{P}_u^{(m)\hspace{-0.25ex}}\hspace{-0.15ex} \hat{d}_{{\hat{k}}}^{(m)\hspace{-0.5ex}}}_{\text{Total MU-interference}} $} \vspace{-1ex} \\ 
		& + \underbrace{n_{D_{k}} + n_{2_{k}}^{(m)}}_{\text{Total Noise}}, \raisetag{1.5\baselineskip} 
	\end{split}	
\end{equation}
where $n_{D_{k}}$ $\sim$ $\mathcal{CN} (\mathbf{0}, \sigma_{n_D}^2)$ and $n_{2_{k}}^{(m)}$ $\sim$ $\mathcal{CN} (\mathbf{0}, \sigma_{n_2}^2)$ are the additive circular symmetric Gaussian noise at $k^{th}$ IoT node. The power constraint of the beamforming matrices can be expressed as $\lVert \mathbf{F}_b\mathbf{B}_b \mathbf{P}_b \rVert^2_F$ = $P_T$ and $\lVert \mathbf{F}_{u,t}^{(m)}\mathbf{B}_{u,t}^{(m)} \mathbf{P}_u^{(m)} \rVert^2_F$ = $P_u^{(m)}$, where $P_T$ and $P_u^{(m)}$ denote the total transmit power of BS and $m^{th}$ UAV, respectively. Then, the achievable rate of first link (i.e., BS $\rightarrow$ $m^{th}$ UAV) is given as follows: 
\begin{equation}
	\mbox{$\squeezespaces{0.05}	\mathrm{R}_1^{\hspace{-0.4ex}(m)} \hspace{-0.35ex}=\hspace{-0.1ex} \log_2 \hspace{-0.25ex}\left|\hspace{-0.15ex}\mathbf{I}_{K} \hspace{-0.15ex} +  \hspace{-0.15ex}\mathbf{Q}_1^{\mathrm{-1}(m)}\hspace{-0.15ex}\mathbf{B}_{u,r}^{(m)}\mathbfcal{H}_1^{(m)} \mathbf{B}_b \mathbf{B}_b^H \mathbfcal{H}_1^{H(m)} \mathbf{B}_{u,r}^{H(m)} \hspace{-0.1ex}\right|$}, 
\end{equation} 
where \mbox{$\squeezespaces{0.05}\mathbf{Q}_1^{\mathrm{-1}(m)} = (\sigma_{n_u}^2 \mathbf{B}_{u,r}^{(m)} \mathbf{F}_{u,r}^{(m)})^\mathrm{-1}  \mathbf{F}_{u,r}^{H(m)} \mathbf{B}_{u,r}^{H(m)}$} and $\mathbfcal{H}_1^{(m)} \hspace{-1ex}=\hspace{-1ex} \mathbf{F}_{u,r}^{(m)}\mathbf{H}_1^{(m)}\mathbf{F}_b$. The signal-to-interference-plus-noise ratio (SINR) of $k^{th}$ IoT node via $m^{th}$ UAV is given as \cite{Coop_Relaying_2014}:
\begin{equation}
	\mbox{$\squeezespaces{0.5}	\gamma_k^{(m)} \hspace{-0.95ex}=\hspace{-0.5ex} \frac{p_{b_{k}}|\mathbf{h}_{D_{k}}^{H} \mathbf{F}_{b}\mathbf{b}_{b_{k}}|^2}{\hspace{-0.25ex}\sum\limits_{\hat{k} \neq k} \hspace{-0.25ex} p_{b_{{\hat{k}}}}\hspace{-0.3ex}|\hspace{-0.3ex}\mathbf{h}_{D_{k}\hspace{-0.25ex}}^{H}\hspace{-0.2ex} \mathbf{F}_{b}\hspace{-0.25ex}\mathbf{b}_{b_{{\hat{k}}}\hspace{-0.4ex}}|^2 \hspace{-0.35ex} + \hspace{-0.25ex}\sigma_{n_D\hspace{-0.35ex}}^2} \hspace{-0.3ex}+\hspace{-0.3ex} \frac{p_{u_{k}}^{(m)}|\mathbf{h}_{2_{k}}^{H(m)} \mathbf{F}_{u,t}^{(m)}\mathbf{b}_{u,t_{k}}^{(m)}|^2}{\hspace{-0.5ex}\sum\limits_{\hat{k} \neq k}\hspace{-0.25ex} p_{u_{{\hat{k}}}}^{(m)} \hspace{-0.25ex}|\hspace{-0.25ex}\mathbf{h}_{2_{k}}^{H(m)}\hspace{-0.25ex} \mathbf{F}_{u,t}^{(m)}\hspace{-0.25ex}\mathbf{b}_{u,t_{{\hat{k}}}\hspace{-0.2ex}}^{(m)}\hspace{-0.25ex}|^2 \hspace{-0.25ex}+ \hspace{-0.25ex} \sigma_{n_2}^{2(m)}}.$} \label{eq:sinr}	
\end{equation} 
The achievable rate for the $m^{th}$ second link (i.e., from $m^{th}$ UAV to $K_m$ users) can be written as: 
\begin{equation}
	\mbox{$\squeezespaces{0.2}	\mathrm{R}_2^{(m)} \hspace{-0.2ex}= \hspace{-0.2ex}\EX \bigl\{\hspace{-0.2ex} \sum_{i = 1}^{K_m}\log_2 \hspace{-0.2ex}(1 + \gamma_i^{(m)}\hspace{-0.2ex})\hspace{-0.2ex}\bigr\}.$} \label{eq:R_sum} 
\end{equation}
For multiple dual-hop MU-mMIMO IoT system, where each $m^{th}$ UAV is deployed at a fixed height $z_u^{m}$, and relaying data to $K_m$ IoT nodes, the total achievable rate can be maximized by the joint optimization of $\mathbf{F}_b$, $\mathbf{B}_b$, $\mathbf{F}_{u,t}^{(m)}$, $\mathbf{F}_{u,r}^{(m)}$, $\mathbf{B}_{u,t}^{(m)}$, $\mathbf{B}_{u,r}^{(m)}$, $\mathbf{P}_b$, $\mathbf{P}_u^{(m)}$ and the UAV location $\mathbf{x}^{(m)} = [x_o^{(m)}, y_o^{(m)}]^T \in \mathbb{R}^{2}$, which is to be optimized within the given deployment area. Then, we can formulate the optimization problem as:
\begin{equation}
	\begin{split}
		&\max_{\left\{\hspace{-0.2ex} \mathbf{F}_b, \hspace{-0.2ex}\mathbf{B}_b,\hspace{-0.2ex} \mathbf{F}_{u,t}^{\hspace{-0.2ex}(m)},\hspace{-0.2ex} \mathbf{B}_{u,t}^{\hspace{-0.2ex}(m)}, \hspace{-0.2ex}\mathbf{F}_{u,r}^{\hspace{-0.2ex}(m)},\hspace{-0.2ex} \mathbf{B}_{u,r}^{\hspace{-0.2ex}(m)},\hspace{-0.2ex} \mathbf{P}_b,\hspace{-0.2ex} \mathbf{P}_u^{\hspace{-0.2ex}(m)},\hspace{-0.2ex} \mathbf{x}^{\hspace{-0.2ex}(m)}\hspace{-0.6ex}\right\}} \quad \hspace{-2ex} \mathrm{R}_T\hspace{-0.2ex} \\
		 &\hspace{4ex}\textrm{s.t.} \hspace{0.7em}C_1\hspace{-0.5ex}: \hspace{-0.5ex}  |\mathbf{F}_{u,t}^{\hspace{-0.2ex}(m)} (i,j)|\hspace{-0.75ex} = \hspace{-0.75ex} \frac{1}{\sqrt{\hspace{-0.5ex}N_{t}}}, |\mathbf{F}_{u,r}^{\hspace{-0.2ex}(m)} (i,j)| \hspace{-0.75ex}= \hspace{-0.75ex} \frac{1}{\sqrt{\hspace{-0.5ex}N_{r}}}, \hspace{0.5ex} \forall i,j,m, \\
		 & \quad \hspace{6ex}C_2\hspace{-0.5ex}: \hspace{-0.5ex}  |\mathbf{F}_{b} (i,j)| = \frac{1}{\sqrt{N_b}}, \forall i,j, \\ 
		 &\quad \hspace{6ex}C_3\hspace{-0.5ex}: \hspace{-0.5ex} \bigcup_{m \in M} K_m = K, \hspace{2ex}\forall m, \\
		&\quad \hspace{6ex}C_4\hspace{-0.5ex}: \hspace{-0.5ex} \EX \{ \left\lVert \mathbf{s}_1 \right\rVert_2^2\} \leq P_T, \EX \{ || \hat{\mathbf{s}}^{(m)} ||_2^2\} \leq P_T, \forall m,  \\
		&\quad \hspace{6ex}C_5\hspace{-0.5ex}: \hspace{-0.5ex} p_{b_k} \geq 0, p_{u_k}^{(m)} \geq 0, \hspace{2ex} \forall k,m, \\ 
		&\quad \hspace{6ex}C_6\hspace{-0.5ex}: \hspace{-0.5ex} \mathbf{x}_\mathrm{min} \leq \mathbf{x}_o^{(m)} \leq \mathbf{x}_\mathrm{max}, \hspace{2ex}\forall m,		
	\end{split} \label{eq:optimization}  \raisetag{1\baselineskip} 
\end{equation} 
where $\mathrm{R}_T = \sum\nolimits_{m=1}^{M}\hspace{-0.4ex} (1/2)\hspace{-0.6ex} \min( \mathrm{R}_1^{\hspace{-0.3ex}(m)}\hspace{-0.2ex}, \mathrm{R}_2^{\hspace{-0.2ex}(m)})$ is the total achievable rate, $C_1$ and $C_2$ refers to the CM constraint due to the use of PSs for UAV and BS, respectively, $C_3$ ensures that the total user count is consistent with the counts for each UAV, $C_4$ indicates the total transmit power constraint for UAV and BS, $C_5$ represents the non-negative allocated power to each IoT node from BS and each UAV, and $C_6$ implies UAV deployment within the given flying span. Here, $[\mathbf{x}_\mathrm{min}, \mathbf{x}_\mathrm{max}] = [(x_\mathrm{min}, y_\mathrm{min}), (x_\mathrm{max}, y_\mathrm{max})]$ represents the deployment range for each UAV in 2-D space. The optimization problem defined in (\ref{eq:optimization}) is non-convex and intractable. Thus, we develop a sub-optimal solution for (\ref{eq:optimization}) in Section III.
\vspace{-1em}
\subsection{Channel Model}
\vspace{-1ex}
We consider mmWave channels for both links. The channel between BS and $m^{th}$ UAV is modeled based on the Saleh-Valenzuela channel model, and is given as: \vspace{-0.5ex}
\begin{equation}
		\mathbf{H}_1^{\hspace{-0.25ex}(m)\hspace{-0.25ex}}\hspace{-0.75ex} =\hspace{-0.95ex} \sum\limits_{c = 1}^{C}\hspace{-0.45ex} \sum\limits_{l = 1}^{L}\hspace{-0.7ex} z_{\hspace{-0.25ex}1_{cl}\hspace{-0.25ex}}^{\hspace{-0.25ex}(m)} \tau_{\hspace{-0.25ex}1_{cl}\hspace{-0.25ex}}^{\hspace{-0.25ex}-\eta(m)} \mathbf{a}_{\hspace{-0.25ex}1,r}^{\hspace{-0.25ex}(m)}\hspace{-0.45ex} ( \hspace{-0.25ex}{{\theta^{\hspace{-0.15ex}(m)}_{\hspace{-0.25ex}r_{cl}\hspace{-0.25ex}}}}\hspace{-0.7ex},{{\phi^{\hspace{-0.25ex}(m)}_{r_{\hspace{-0.25ex}cl}\hspace{-0.25ex}}}}\hspace{-0.25ex}) \mathbf{a}_{\hspace{-0.25ex}1,t}^{\hspace{-0.25ex}T(m)}\hspace{-0.3ex} (\hspace{-0.25ex} {{\theta_{\hspace{-0.25ex}t_{cl}}^{\hspace{-0.25ex}(m)}}}\hspace{-0.7ex},{{\phi_{\hspace{-0.25ex}t_{cl}}^{\hspace{-0.25ex}(m)}}}\hspace{-0.25ex} )\hspace{-0.1ex},
\end{equation}
where $C$ is the total number of clusters, $L$ is the total number of paths, $\eta$ is the path loss exponent, $z_{1_{cl}}^{(m)}$ is the complex gain of $l^{th}$ path in $c^{th}$ cluster for $m^{th}$ UAV, and ${\bf{a}}_{1,j}^{(m)}(\cdot,\cdot)$ is the corresponding transmit or receive array steering vector for uniform rectangular array (URA), which is given as \cite{mahmood2023deep}: 
\begin{equation}
	\begin{split}
		{\bf{a}}_{1,j}^{(m)}( {{\theta^{(m)}, \phi^{(m)}}}) &= \big[ {1, \cdots,{e^{ -j2\pi d\left( {N_x - 1} \right) {{\sin (\theta^{(m)}) \cos (\phi^{(m)})}} }}} \big]  \\
		&\hspace{-1em} \otimes\big[{1,\cdots,{e^{ -j2\pi d\left( {N_y - 1} \right) {{\sin (\theta^{(m)}) \sin (\phi^{(m)})}} }}} \big], \raisetag{1\baselineskip}
	\end{split}  \label{eq_phase_vector} 
\end{equation} 
where $j = \{t,r\}$, $d$ is the inter-element spacing, and $N_x(N_y)$ is the horizontal (vertical) size of corresponding antenna array at BS and UAV.  Here, the angles ${\theta_{j_{cl}}^{(m)}} \in \big[ {{\theta}_{j_c}^{(m)} - {\delta_{j_c}^{^{\theta (m)}}}}, {{\theta}_{j_c}^{(m)} + {\delta_{j_c}^{^{\theta (m)}}}} \big]$ and   
${\phi _{j_{cl}}^{(m)}} \in \big[ {{\phi}_{j_c}^{(m)} - {\delta_{j_c}^{\phi (m)}}}, {{\phi}_{j_c}^{(m)} + {\delta_{j_c}^{\phi (m)}}} \big]$ are the elevation and azimuth AoD($j=t$) or AoA ($j=r$) for $l^{th}$ path in channel $\mathbf{H}_1^{(m)}$, respectively. Here, ${\theta}_{j_c}^{(m)}$ and ${\phi}_{j_c}^{(m)}$ are the mean elevation and azimuth angles, respectively with ${\delta_{j_c}^{^{\theta (m)}}}$ (${\delta_{j_c}^{\phi (m)}}$) represents the elevation(azimuth) angle spread. The channel vector between the UAV(or BS) and the $k^{th}$ IoT node can be written as:
\begin{equation}
		\mathbf{h}_{i,k}^T \hspace{-0.3ex} =\hspace{-0.3ex}  \sum\nolimits_{q = 1}^{Q} z_{i,k_q}\tau_{i,k_q}^{-\eta}  \mathbf{a}(\theta_{k_q}, \phi_{k_q})\hspace{-0.3ex} = \hspace{-0.3ex} \mathbf{z}_{i,k}^T \mathbf{A}_{i,k} \in \mathbb{C}^{N}, \label{eq:second_channel} 
\end{equation}
where $i = \{D,2\}$, $Q$ is the total number of downlink paths from UAV(or BS) to IoT nodes, $z_{i,k_{q}}$ $\sim$ $\mathcal{C N}(0, \frac{1}{Q})$ is the complex path gain of $q^{th}$ path, $\mathbf{a}(\cdot, \cdot)$ $\in$ $\mathbb{C}^{N}$ is the downlink array phase response vector. Then, the complete channel matrix for $K$ IoT nodes can be written as:
\begin{equation}
	\mathbf{H}_{i}=[\mathbf{h}_{i, 1}, \cdots, \mathbf{h}_{i, K}]^{T}=\mathbf{Z}_{i} \mathbf{A}_i \in \mathbb{C}^{K \times N_{t}},
\end{equation}
where $\mathbf{Z}_{i}=[\mathbf{z}_{i, 1}, \cdots, \mathbf{z}_{i, K}]^{T} \in \mathbb{C}^{K \times Q}$ is the complete path gain matrix for all downlink IoT nodes and $\mathbf{A}_{i,k} \in \mathbb{C}^{Q \times N}$ is the slow time-varying array phase response matrix.
\vspace{-1ex}
\section{Proposed Joint User Association, Multiple UAV Positioning, PA \& Hybrid Beamforming}
In this section, our objective is to optimize each UAV location jointly with PA from BS and UAV, and sequentially design HBF stages for BS and each UAV to reduce the channel state information (CSI) overhead size while maximizing the throughput of a multiple UAV-assisted MU-mMIMO IoT system. First, we discuss the UAV-user association using K-means-based user clustering.
\subsection{UAV-User Association}
The proposed scheme leverages K-means-based user association to assign $K_m$ users to nearest $m^{th}$ UAV while maintaining exclusive user-UAV associations. The objective is to minimize the sum of squared distances between users and their assigned $m^{th}$ UAV (i.e., $\min \sum_{m=1}^{M} \sum_{k=1}^{K} z_{km} ||\mathbf{x}_k - \mathbf{x}_u^{(m)}||^2$). Here, $z_{km}$ is the assignment variable, which is defined as:
\begin{equation}
	z_{km} = \begin{cases}
		1, & \text{if user $k$ is assigned to UAV $m$}, \\
		0, & \text{otherwise}.
	\end{cases}
\end{equation} 
\subsection{Joint UAV Deployment, Optimal PA and HBF Design}
In this section, our objective is to design the HBF stages for BS and $M$ UAVs by using sequential optimization. Initially, both RF and BB stages are constructed using some fixed UAV locations. Then, we employ swarm intelligence to optimize each UAV location as well as PA from BS and each UAV for maximum total achievable rate. Finally, the RF and BB stages are re-formulated for the optimal UAV location as well as the allocated power in the MU PA blocks $\mathbf{P}_b$ and $\mathbf{P}_u^{(m)}$.
\vspace{-1ex}
\subsubsection{RF \& BB Stage Design}
The RF beamforming stage for BS and each UAV (both transmit and receive) are designed as:
\begin{equation}\label{eq:eq_TX_RF_1}  \vspace{-0.5ex}
	{{\bf{F}}} \hspace{-0.75ex}=\hspace{-0.75ex} \big[ \hspace{-0.25ex}{{\bf{e}}_j\big(\hspace{-0.35ex} {\lambda ^{u_1}_{x}\hspace{-0.3ex},\lambda ^{k_1}_{y}} \hspace{-0.5ex}\big), \hspace{-0.25ex} \cdots,\hspace{-0.25ex} {\bf{e}}_j\big(\hspace{-0.25ex} {\lambda ^{u_{N_{{RF}}}}_{x} \hspace{-0.95ex},\lambda ^{k_{N_{{RF}}}}_{y}} \hspace{-0.5ex}\big)}\hspace{-0.25ex} \big]\hspace{-0.75ex} \in \hspace{-0.5ex}\mathbb{C}^{N_{T} \times N_{{RF}}},\hspace{-0.25ex}
\end{equation}
where $j$ $=$ $\{t,r\}$ and $\mathbf{e}(\cdot,\cdot)$ is the corresponding transmit or receive steering vector, which is defined as ${\bf{e}}\hspace{-0.5ex}\left( {{\theta, \phi}} \right) \hspace{-0.5ex} =\hspace{-0.75ex} \frac{1}{\mathcal{N}_t}\big[ {1,{e^{j2\pi d  {{\sin (\theta) \cos (\phi)}} }}, \cdots,{e^{j2\pi d\left( {\mathcal{N}_{x,t} - 1} \right) {{\sin (\theta) \cos (\phi)}} }}} \big]^T \otimes \hspace{-0.5ex}\big[ {1,{e^{j2\pi d  {{\sin (\theta) \sin (\phi)}} }}, \cdots,{e^{j2\pi d\left( {\mathcal{N}_{y,t} - 1} \right) {{\sin (\theta) \sin (\phi)}} }}} \big]^T$, where $\mathcal{N}_T = \{N_b,N_t, N_r\}$. Here, the RF beamformers are constructed via quantized angle-pairs, which are defined as ${{\lambda _{x}^u} \hspace{-0.5ex}=\hspace{-0.5ex} -1 + \frac{2u-1}{{{\mathcal{N}_{x,t}}}}}$ for $u = 1, \cdots,{\mathcal{N}_{x,t}}$ and ${{\lambda _{y}^k} = -1 + \frac{2k-1}{{{\mathcal{N}_{y,t}}}}}$ for $k = 1, \cdots,{\mathcal{N}_{y,t}}$. The quantized angle-pairs reduces the number of RF chains at BS and each UAV while providing complete AoD/AoA supports, which are defined as: \vspace{-0.5ex}
\begin{align}
	\textrm{AoD} = \big\lbrace {\sin \left( \theta  \right){{\left[ {\cos \left( \phi  \right),\sin \left( \phi \right)} \right]}}} \big| {\theta} \in \bm{\theta}_t, {\phi} \in \bm{\phi}_t \big\rbrace	,\label{eq_AoD_Supp} \\
	\textrm{AoA} = \big\lbrace {\sin \left( \theta  \right){{\left[ {\cos \left( \phi  \right),\sin \left( \phi \right)} \right]}}} \big| 	\theta \in{\bm {\theta }}_r, \phi \in{\bm {\phi }}_r\big\rbrace, \label{eq_AoA_Supp} 
\end{align}
where $\bm{\theta}_i = \big[ {{\theta _{i}} - {\delta_{i}^\theta}}, {{\theta _{i}} + {\delta _{i}^\theta}} \big]$ and $\bm{\phi}_i = \big[ {{\phi _{i}} - {\delta_{i}^\phi}}, {{\phi _{i}} + {\delta _{i}^\phi}} \big]$ denote the azimuth and elevation angle supports, respectively. After designing the transmit and receive RF beamformers for BS, and each UAV, the effective channel matrices $\mathbfcal{H}_1^{(m)}$ and $\mathbfcal{H}_2^{(m)}$ as seen from the BB-stages are given as follows:
\begin{equation} 
	\mathbfcal{H}_1^{(m)} = \mathbf{F}_{u,r}^{(m)}\mathbf{H}_1^{(m)}\mathbf{F}_b = \mathbf{U}_1^{(m)} \mathbf{\Sigma}_1^{(m)} \mathbf{V}_1^{H(m)}, \label{eq:reduced_channel_1}
	\end{equation}
\begin{align}
	\mathbfcal{H}_{2}^{(m)}\hspace{-1ex}=\hspace{-1ex}\left[\hspace{-1ex}\begin{array}{ccc} \hspace{-0.7ex}
	\mathbf{H}_{2,1}^{(m)} \mathbf{F}_{u,t,1}^{(m)} & \ldots & \mathbf{H}_{2,1}^{(m)} \mathbf{F}_{u,t,G}^{(m)} \\
	\vdots & \ddots & \vdots \\
	\mathbf{H}_{2,G}^{(m)} \mathbf{F}_{u,t,1}^{(m)} & \ldots & \mathbf{H}_{2,G}^{(m)} \mathbf{F}_{u,t,G}^{(m)}
\end{array}\hspace{-1ex}\right]\hspace{-1ex}, \hspace{-0.7ex}
%	\mathbfcal{H}_{\hspace{-0.3ex}D}\hspace{-0.95ex}=\hspace{-0.9ex}\left[\hspace{-1ex}\begin{array}{ccc} \hspace{-0.7ex}
%	\mathbf{H}_{\hspace{-0.2ex}D,1\hspace{-0.4ex}} \mathbf{F}_{\hspace{-0.2ex}b,1\hspace{-0.4ex}} \hspace{-2ex}&\hspace{-2.5ex} \ldots \hspace{-4ex}&\hspace{-2ex} \mathbf{H}_{\hspace{-0.2ex}D,1\hspace{-0.4ex}} \mathbf{F}_{\hspace{-0.2ex}b,G\hspace{-0.4ex}} \hspace{-2ex} \\
%	\vdots \hspace{-1ex}&\hspace{-2.5ex} \ddots \hspace{-4ex}&\hspace{-1ex} \vdots\hspace{-2ex} \\
%	\hspace{-0.7ex}\mathbf{H}_{\hspace{-0.2ex}D\hspace{-0.1ex},G\hspace{-0.4ex}} \mathbf{F}_{\hspace{-0.2ex}b\hspace{-0.1ex},1\hspace{-0.4ex}} \hspace{-2ex}&\hspace{-2ex} \ldots \hspace{-4ex}&\hspace{-2ex} \mathbf{H}_{\hspace{-0.2ex}D\hspace{-0.1ex},G\hspace{-0.3ex}} \mathbf{F}_{\hspace{-0.2ex}b\hspace{-0.1ex},G\hspace{-0.3ex}}\hspace{-2ex}
%\end{array}\hspace{-1.2ex}\right]\hspace{-1ex},
 \label{eq:effective_channel_2}  \vspace{-0.5ex} 
\end{align}
where $\mathbf{U}_i^{(m)}$ $\in$ $\mathbb{C}^{N_{{RF}_u} \times rank(\mathbfcal{H}_1^{(m)})}$ and $\mathbf{V}_i^{(m)}$ $\in$ $\mathbb{C}^{N_{{RF}_b} \times rank(\mathbfcal{H}_1^{(m)})}$ are tall unitary matrices and $\mathbf{\Sigma}_1^{(m)}$ is the diagonal matrix with singular values in the decreasing order for $m^{th}$ UAV. Then, $\mathbf{B}_{u,r}^{(m)}$ for $m^{th}$ UAV is defined as:
\begin{equation}
\mathbf{B}_{u,r}^{(m)} = \mathbf{U}_1^{H(m)}. \label{eq:BB_1st_link}
\end{equation}
The reduced-size effective CSI $\bm{\mathcal{H}}_2^{(m)}$ given in (\ref{eq:effective_channel_2}) is employed for designing $\mathbf{B}_{u,t}^{(m)}$ by using regularized zero-forcing (RZF) technique, and is defined as follows: 	
\begin{equation}
	\mathbf{B}_{u,t}^{(m)} \hspace{-0.75ex} = \hspace{-0.75ex}(\hspace{-0.35ex}\mathbfcal{H}_2^{\hspace{-0.25ex}H(m)\hspace{-0.25ex}} \mathbfcal{H}_2^{\hspace{-0.25ex}(m)\hspace{-0.25ex}} \hspace{-0.25ex}+\hspace{-0.25ex}\beta^{\hspace{-0.25ex}(m)\hspace{-0.25ex}} N_{\hspace{-0.25ex}{RF}_u\hspace{-0.25ex}}\mathbf{I}_{\hspace{-0.25ex}N_{{RF}_u}\hspace{-0.25ex}}\hspace{-0.35ex})^{\hspace{-0.25ex}-1\hspace{-0.25ex}}\mathbfcal{H}_2^{\hspace{-0.25ex}H(m)\hspace{-0.25ex}}, \label{eq:BB_precoder}  \vspace{-1ex}
\end{equation}
where $\beta^{(m)} =\frac{\sigma^{2(m)}}{P_T^{(m)}}$ is the regularization parameter and \squeezespaces{0.1}$\mathbf{I}_{N_{{RF}_u}}$ $\in$ $\mathbb{C}^{N_{{RF}_u} \times N_{{RF}_u}}$. The optimal design of $\mathbf{B}_{b}$ is formulated using an effective channel $\mathbfcal{H}_{1D}$, which constitutes all channel components from BS (i.e., $\mathbf{H}_D$ and $\mathbf{H}_2^{(m)}, \forall m=1,\cdots, M$). Then, the total effective channel can be written as: 
\begin{equation}
	\mathbfcal{H}_{1D} = [\mathbf{H}_D, \mathbf{H}_1^{(1)}, \mathbf{H}_1^{(2)},\cdots, \mathbf{H}_1^{(M)}] \in \mathbb{C}^{(M N_r + K)\times K}.
\end{equation}
By using SVD of effective channel $\mathbfcal{H}_{1D}$, we can design $\mathbf{B}_b$ by using tall unitary matrix $\mathbf{V}_{1D}$ $\in$ $\mathbb{C}^{N_{{RF}_b} \times rank(\mathbfcal{H}_{1D})}$ as \cite{mahmood2023deep}: 
\begin{equation} \label{eq:BB_BS}
	\mathbf{B}_b = \sqrt{\frac{P_T}{K}}\mathbf{V}_{1D} \in \mathbb{C}^{N_{{RF}_b} \times K}. \vspace{-0.5ex}
\end{equation}
 \begin{table}[!t]
	\centering
	\caption{Simulation Parameters \cite{koc2020Access}\vspace{-2ex}}
	\resizebox{\columnwidth}{!}{
		\begin{tabular}{|c|c|c|c|}
			\hline
			\multicolumn{2}{|c|}{Number of antennas}	& \multicolumn{2}{c|}{$(N_b, N_r, N_t) = 64$} \\ \hline 	
			Number of paths & Path loss exponent & $L = 10$ &  $3.6$ \\ \hline 	
			Frequency &  Channel Bandwidth & 28 GHz & 100 MHz  \\ \hline
			Noise PSD & Number of UAVs $M$ & $-174$ dBm/Hz & 2 or 3 \\ \hline 	
			BS &  UAV height & 10 m & 20 m  \\ \hline			
			UAV x-axis range & UAV y-axis range & $[x_{min}, x_{max}] = \left[0,100\right] m$ & $[y_{min}, y_{max}] = \left[0,100\right] m$ \\ \hline
			Azimuth AoD/AoA ($1^{st}$ link) & Azimuth AoD/AoA ($2^{nd}$ link) & $120^{\circ}$  & $150^{\circ}$  \\ \hline
			Elevation AoD/AoA ($1^{st}$ link) & Elevation AoD/AoA ($2^{nd}$ link) & $60^{\circ}$  & $30^{\circ}$  \\ \hline  	
			Azimuth/Elevation Angle Spread & \# of network realizations & $\pm 10^{\circ}$ & 2000 \\ \hline		
		\end{tabular} 
	} 
\end{table}
\vspace{-2ex}
\subsubsection{Joint Multiple UAV Positioning and Optimal PA}
After the design of RF and BB stages for BS and UAVs, the optimization problem given in (\ref{eq:optimization}) can be reformulated as:  
\begin{equation}
	\begin{aligned}
		&\max_{\left\{\mathbf{P}_b,\mathbf{P}_u^{(m)},\mathbf{x}^{(m)}\right\}} \quad \hspace{-1ex} \mathrm{R}_T \\
		& \hspace{5ex}\textrm{s.t.} \hspace{5ex} C_4 - C_6. 
	\end{aligned} \label{eq:optimization_reformulated} 
\end{equation} 
This resulting problem in (\ref{eq:optimization_reformulated}) is still non-convex due to the joint dependence of both the allocated powers $p_{b_k}$, $p_{u_k}^{(m)}$ and the UAV location $\mathbf{x}^{(m)} = [x_o^{(m)}, y_o^{(m)}]^T$ on the SINR expression in (\ref{eq:sinr}), which is used in the sum-rate $\mathrm{R}_2^{(m)}$ calculation as given in (\ref{eq:R_sum}). To overcome this challenge, we propose sequential optimization using swarm intelligence, which employs multiple agents, called particles, to explore the search space of objective function given in (\ref{eq:optimization_reformulated}). Initially, $N_p$ particles are randomly placed in search space, where each particle communicates with other particles to share their personal best and update the current global best solution for the objective function. The particles then move iteratively for $T$ iterations to reach the global optimum solution. In particular, each UAV location $\mathbf{x}_u^{(m)} = [x_u^{(m)}, y_u^{(m)}]^T$ and $\mathbf{P}_b$, $\mathbf{P}_u^{(m)}$ are optimized by using particle swarm optimization (PSO)-based algorithmic solution while maximizing the total achievable rate. Here, the $i^{th}$ particle at the $t^{th}$ iteration now represents an instance of the each UAV location and multi-user PA matrices, which is given as follows:
\begin{equation}
	\begin{split}
	&\mbox{$\squeezespaces{0.1}	\mathbf{J}_{\hspace{-0.25ex}p_i\hspace{-0.25ex}}^{\hspace{-0.25ex}(t)\hspace{-0.25ex}} \hspace{-0.25ex}=\hspace{-0.25ex} [\hspace{-0.25ex}\mathbf{X}_i^{\hspace{-0.25ex}(t)}, \hat{\mathbf{P}}_{\hspace{-0.25ex}b_i}^{(t)},\hat{\mathbf{P}}_{\hspace{-0.25ex}u_{i}}^{(t)}\hspace{-0.25ex}]^T \hspace{-0.55ex}=\hspace{-0.25ex} [x_{1_i}^{(t)}\hspace{-0.25ex}, y_{1_i}^{(t)}\hspace{-0.25ex},\cdots\hspace{-0.25ex}, x_{M_i}^{(t)}\hspace{-0.25ex}, y_{M_i}^{(t)}\hspace{-0.25ex},\sqrt{\hat{p}_{b,1_i}^{(t)}}\hspace{-0.5ex},\cdots,\hspace{-0.5ex} \sqrt{\hat{p}_{b,K_i}^{(t)}}, \cdots, $} \\
	 &\mbox{$\squeezespaces{0.1} \cdots,\sqrt{\hspace{-0.25ex}\hat{p}_{\hspace{-0.25ex}u_1,1_i\hspace{-0.25ex}}^{\hspace{-0.25ex}(t)\hspace{-0.25ex}}}\hspace{-0.5ex},\cdots,\hspace{-0.5ex} \sqrt{\hspace{-0.25ex}\hat{p}_{\hspace{-0.25ex}u_1,K_i\hspace{-0.25ex}}^{\hspace{-0.25ex}(t)}\hspace{-0.25ex}}, \cdots,\sqrt{\hspace{-0.25ex}\hat{p}_{\hspace{-0.25ex}u_M,1_i\hspace{-0.15ex}}^{\hspace{-0.25ex}(t)}\hspace{-0.25ex}}\hspace{-0.15ex},\cdots,\hspace{-0.5ex} \sqrt{\hspace{-0.25ex}\hat{p}_{\hspace{-0.25ex}u_M,K_i\hspace{-0.25ex}}^{\hspace{-0.25ex}(t)}}]^T \in \mathbb{R}^{\hspace{-0.25ex}2K+2M\hspace{-0.25ex}}$}, 
	\end{split}    \label{eq:Joint_PSO}    \raisetag{1\baselineskip}
\end{equation}
where each particle $i$ represents the $M$ UAV positions and PA to $K$ IoT users from BS and each $m^{th}$ UAV, and calculates the objective function as $R_{\hspace{-0.15ex}T}(\hspace{-0.15ex}\mathbf{F}_b\hspace{-0.15ex}, \mathbf{B}_b\hspace{-0.15ex}, \mathbf{F}_{\hspace{-0.15ex}u,t\hspace{-0.15ex}}^{\hspace{-0.15ex}(m)\hspace{-0.15ex}}\hspace{-0.15ex}, \mathbf{B}_{\hspace{-0.15ex}u,t\hspace{-0.15ex}}^{\hspace{-0.15ex}(m)\hspace{-0.15ex}}\hspace{-0.25ex}, \mathbf{F}_{\hspace{-0.15ex}u,r\hspace{-0.15ex}}^{\hspace{-0.15ex}(m)\hspace{-0.15ex}}\hspace{-0.25ex}, \mathbf{B}_{\hspace{-0.15ex}u,r\hspace{-0.15ex}}^{\hspace{-0.15ex}(m)\hspace{-0.15ex}}\hspace{-0.25ex}, \kappa_{b_i}^{(t)}\hat{\mathbf{P}}_{b_i}^{(t)}\hspace{-0.15ex}, \kappa_{u_i}^{(t)}\hat{\mathbf{P}}_{u_i}^{(t)}\hspace{-0.15ex}, \mathbf{X}_i^{(t)}\hspace{-0.15ex})$. We define $\hat{\mathbf{P}}_{b_i}^{(t)}=\operatorname{diag}(\sqrt{\hat{p}_{b,1_i}^{(t)}}, \cdots, \sqrt{\hat{p}_{b, K_i}^{(t)}}) \in \mathbb{R}^{K \times K}$ and $\hat{\mathbf{P}}_{u_{i}}^{(t)}= [\hat{\mathbf{P}}_{u_{1_i}}^{(t)},\cdots, \hat{\mathbf{P}}_{u_{M_i}}^{(t)}]$, where $\hat{\mathbf{P}}_{u_{m_i}}^{(t)}=\operatorname{diag}(\sqrt{\hat{p}_{u_m,1_i}^{(t)}}, \cdots, \sqrt{\hat{p}_{u_m, K_i}^{(t)}}) \in \mathbb{R}^{K \times K}$ as the normalized PA matrices with $\hat{p}_{b,k_i}^{(t)}$, $\hat{p}_{u_m,k_i}^{(t)} \in [0,1]$. Then, the transmit power constraints for $\hat{\mathbf{P}}_{b_i}^{(t)}$ and $\hat{\mathbf{P}}_{u_{i}}^{(t)}$ are satisfied by defining $\mathbf{P}_{b_i}^{(t)}=\kappa_{b_i}^{(t)} \mathbf{\hat{P}}_{b_i}^{(t)}$ and $\mathbf{P}_{u_i}^{(t)}=\kappa_{u_i}^{(t)} \mathbf{\hat{P}}_{u_i}^{(t)}$. The position $\mathbf{J}_{p_i}^{(t)}$ and velocity $\mathbf{J}_{v_i}^{(t)}$ for $i^{th}$ particle during $t^{th}$ iteration are updated as follows:
\begin{equation}
	\mathbf{J}_{p_i}^{(t+1)}= \mathbf{J}_{p_i}^{(t)} + \mathbf{J}_{v_i}^{(t+1)}, \label{eq:position_PSOLPA} 
\end{equation}
\begin{equation}
	\mbox{$\squeezespaces{0.1} \mathbf{J}_{v_i}^{(t+1)}\hspace{-0.5ex}=\gamma_1\hspace{-0.25ex} \mathbf{Y}_1^{(t)}\hspace{-0.5ex}(\hspace{-0.35ex}\mathbf{J}_{p_\mathrm{best}}^{(t)}\hspace{-0.95ex}-\mathbf{J}_{p_i}^{(t)}\hspace{-0.25ex})\hspace{-0.25ex}+\hspace{-0.35ex}\gamma_2\hspace{-0.25ex} \mathbf{Y}_2^{(t)}\hspace{-0.5ex}(\hspace{-0.35ex}\mathbf{J}_{p_{\mathrm{best}_i}}^{(t)}\hspace{-0.95ex}-\mathbf{J}_{p_i}^{(t)}\hspace{-0.25ex})\hspace{-0.15ex}+\hspace{-0.25ex}\gamma_3^{(t)} \mathbf{J}_{v_i}^{(t)}\hspace{-0.25ex}$}. \label{eq:velocity_PSOLPA}  \vspace{-1ex}
\end{equation}
Finally, the personal and global best solutions for $i^{th}$ particle during $t^{th}$ iteration are obtained as follows: \vspace{-0.5ex}
\begin{equation}
	\begin{split}
&	\mathbf{J}_{p_{\mathrm{best}_i}}^{(t)} = \hspace{-1em} \argmax_{\mathbf{J}_{p_i}^{(t^*)}, \forall t^* = 0,1,\cdots, t} R_T(\mathbf{F}_{b},\mathbf{B}_{b},\mathbf{F}_{u,t}^{(m)},\mathbf{B}_{u,t}^{(m)},\mathbf{F}_{u,r}^{(m)}, \cdots, \\ &\hspace{6.5em}\cdots, \mathbf{B}_{u,r}^{(m)},\kappa_{b_i}^{(t^*)}\hat{\mathbf{P}}_{b_i}^{(t^*)},\kappa_{u_i}^{(t^*)}\hat{\mathbf{P}}_{u_i}^{(t^*)},\mathbf{X}_{i}^{(t^*)}),   \vspace{-0.5ex}
		\end{split}\label{eq: personal_best_PSOLPA} \raisetag{1\baselineskip}
\end{equation} 
\begin{equation} 
	\begin{split}
&	\mathbf{J}_{p_\mathrm{best}}^{(t)} = \hspace{-1ex}\argmax_{\mathbf{J}_{p_{\mathrm{best}_i}}^{(t)}, \forall i = 0,1,\cdots, M_p} R_T(\mathbf{F}_b,\mathbf{B}_b, \mathbf{F}_{u,t}^{(m)},\mathbf{B}_{u,t}^{(m)},\mathbf{F}_{u,r}^{(m)}, \cdots, \\
\vspace{-2ex}& \vspace{-2ex} \hspace{4em}\cdots,\mathbf{B}_{u,r}^{(m)},\kappa_{\mathrm{best},b_i}^{(t)}\hat{\mathbf{P}}_{\mathrm{best,b_i}}^{(t)},\kappa_{\mathrm{best},u_i}^{(t)}\hat{\mathbf{P}}_{\mathrm{best,u_i}}^{(t)},\mathbf{X}_i^{(t)}).
\end{split} \label{eq:global_best_PSOLPA} 	\raisetag{1.9\baselineskip}
\vspace{-5ex}
\end{equation}
After $T$ iterations, we update $\mathbf{x}^{(m)} = \mathbf{X}_{\mathrm{best}}^{(T)}$, $\mathbf{P}_b=\kappa_{\mathrm{best},b}^{(T)} \hat{\mathbf{P}}_{\mathrm{best},b}^{(T)}$ and $\mathbf{P}_u^{(m)}=\kappa_{\mathrm{best},u}^{(T)} \hat{\mathbf{P}}_{\mathrm{best},u}^{(T)}$.
\begin{figure}[!t] 
	% 	\centering
	\subfloat[\label{fig:fig3a}]{% 
		\includegraphics[width =0.485\columnwidth,height=4cm]{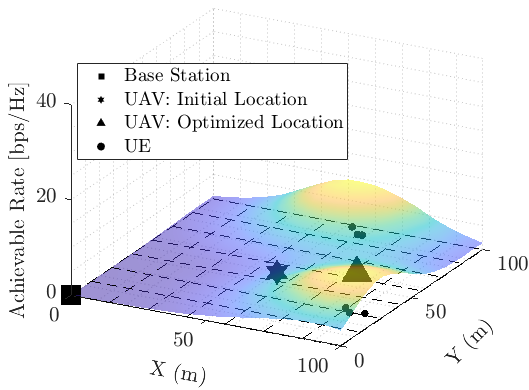} 
	} \hfil 
	\subfloat[\label{fig:fig3b}]{% 
		\includegraphics[width =0.485\columnwidth, height= 4cm]{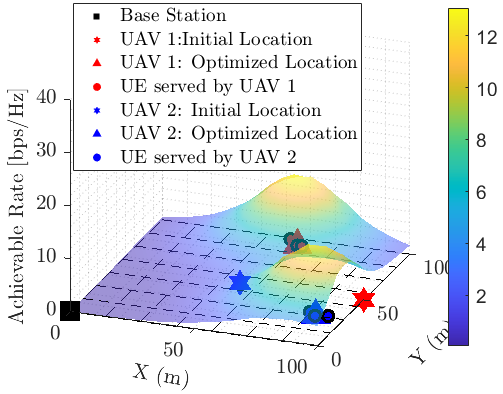} 
	} \vspace{-1ex}
	\caption{Achievable rate $\mathrm{R}_2$ vs. $(x-y)$-coordinates at $P_T^{(m)}$ = 20 dBm. (a) Single UAV deployment ($M=1$). (b) Multiple UAV deployment ($M=2$).}
	\label{fig:fig3}
	\vspace{0.5ex}
\end{figure}
\vspace{-1.2ex}
\section{Illustrative Results}
\vspace{-1ex}
In this section, the Monte-Carlo simulation results are presented based on the proposed scheme. Table I outlines the simulation setup based on the 3D micro-cell scenario \cite{mahmood2023deep} for the results discussed hereafter. The PSO parameters are chosen as: $N_p$$=$$20$, $\gamma_1$$=$$\gamma_2$$=2$ and $\gamma_3$$= 1.1$. In Fig. \ref{fig:fig3}(a), we compare the achievable rate $\mathrm{R}_2$ versus transmit power for the following two cases: 1) a single UAV ($M=1$) deployed at initial fixed location $(x_u,y_u) = (50,50)$; and 2) multiple UAVs ($M=2$) deployed at initial fixed locations $(x_u^{(1)},y_u^{(1)}) = (50,50)$, $(x_u^{(2)},y_u^{(2)}) = (100,50)$. We consider a practical user distribution scenario where the users are located at multiple locations (i.e., ($x_k,y_k$) $\in$ [50,100]) from BS, which is located at $(x_b, y_b) = (0,0)$. It can be seen from Fig. \ref{fig:fig3}(a) that the proposed scheme can optimize the UAV location, however, it can only achieve a sub-optimal solution as a single UAV can not be positioned optimally to support a large number of users. Moreover, each $k^{th}$ user experience interference from $K-1$ users, which leads to low achievable rate. To improve the performance, a multiple UAV-assisted system (i.e., $M=2$) is used in Fig. \ref{fig:fig3}(b), where each UAV can support $K_m = K/M$ users based on the proposed scheme and find the optimal deployments close to its associated $K_m$ users. It can be seen that using the proposed PSO-based UAV location and PA
(J-HBF-PSOLPA) scheme, each UAV can optimally cluster its users, and then find the optimal deployment while achieving the maximum achievable rate. Thus, a multiple UAV-assisted MU-mMIMO system can overcome the coverage and capacity limitation of a single UAV in more practical scenarios. \par 
\begin{figure}[!t]
%	\centering
%	\captionsetup{justification=centering}	
	\includegraphics[height=5.5cm, width=1\columnwidth]{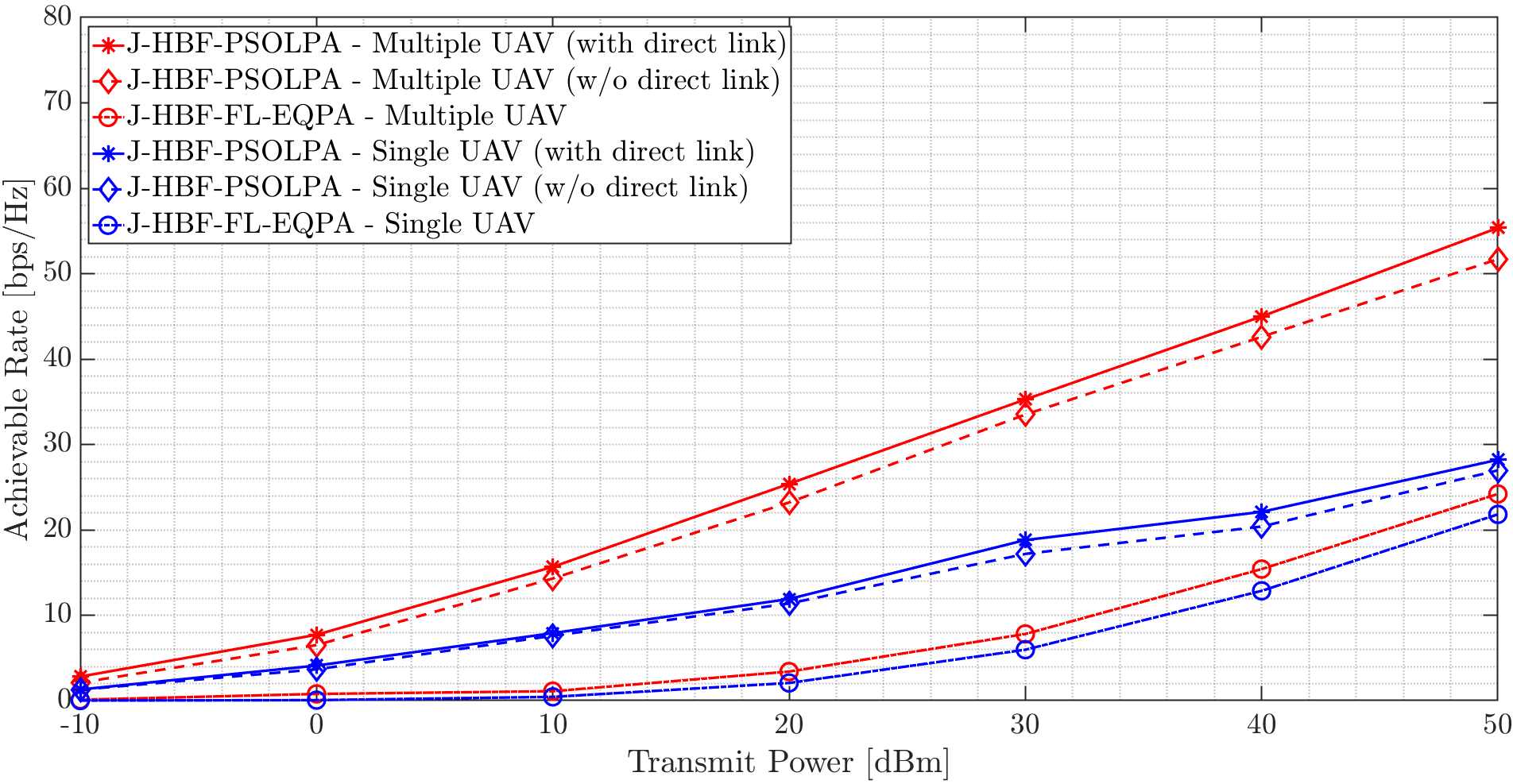} 
	\caption{Total AR $\mathrm{R}_T$ vs. $P_T$ for single and multiple UAV system.}
	\label{fig:fig4}
\end{figure}
\begin{figure}[!t]
	\includegraphics[height=5.5cm, width=1\columnwidth]{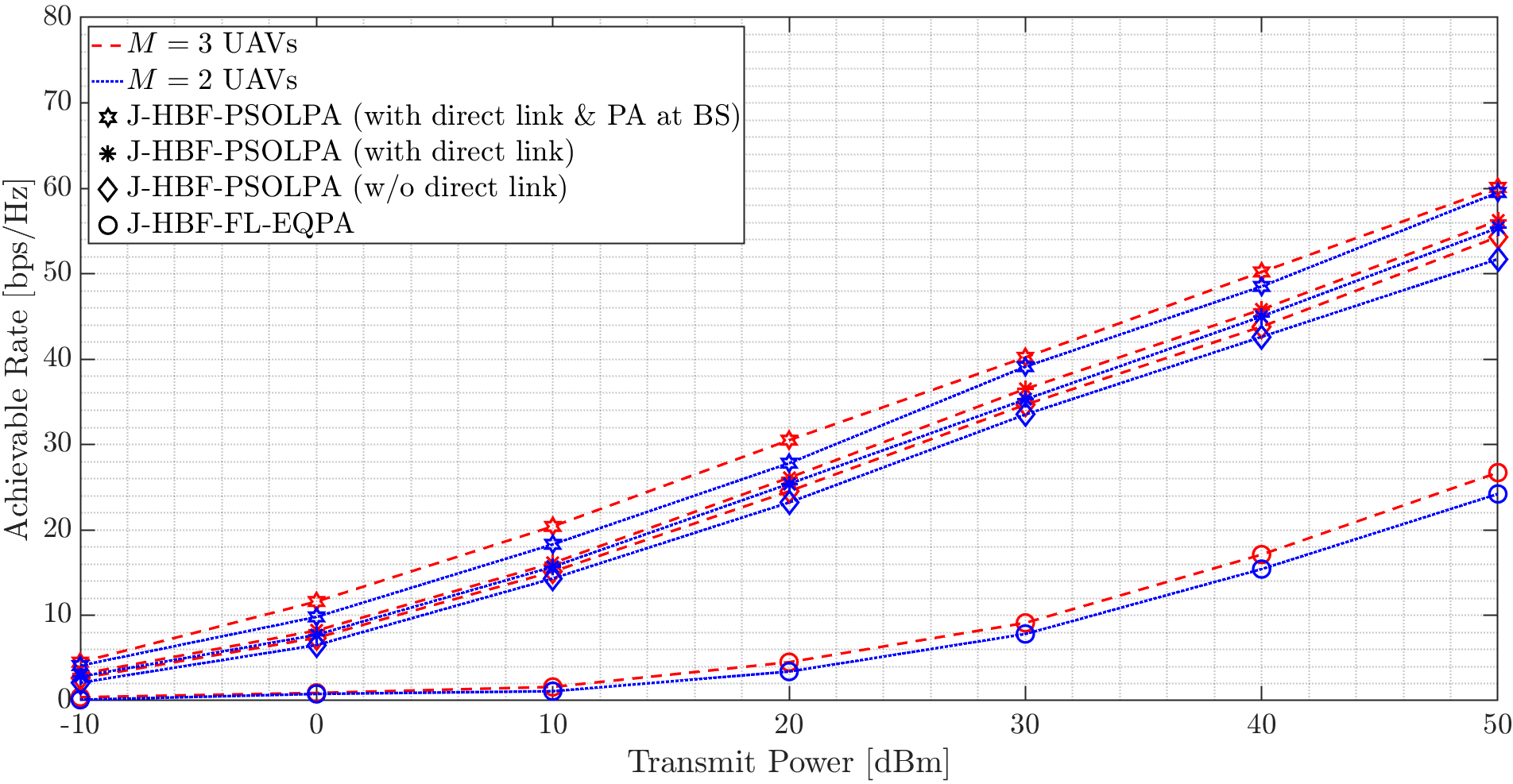} 
	\caption{Total AR $\mathrm{R}_T$ vs. $P_T$ for 2 and 3 UAV system.}
	\label{fig:fig5}
	\vspace{-1ex}
\end{figure}

Fig. \ref{fig:fig4} compares the achievable rate of a multiple UAV system versus a single UAV system for the following cases: 1) J-HBF-PSOLPA with direct link from BS (we consider optimal PA at each UAV while considering equal PA from BS); 2) J-HBF-PSOLPA without direct link from BS; and 3) fixed UAV location and equal PA (FL-EQPA). It can be seen that by using $M=2$ UAV can provide approximately 2.5 times the achievable rate when compared to a single UAV case ($M=1$). For instance, the achievable rate at $P_T = 50$ dBm is increased from 27 bps/Hz to 58 bps/Hz, which indicates around 250$\%$ increase in total achievable rate. Moreover, compared to FL-EQPA, the proposed scheme can significantly enhance the performance by optimizing the UAV locations and the allocated powers. The inclusion of direct link from BS to users (i.e., cooperative relaying) can provide an additional 4 to 5 bps/Hz rate improvement for all cases. Fig. \ref{fig:fig5} compares the achievable rate for different number of UAVs (i.e., for $M=2$ or $M=3$) for the following four cases: 1) J-HBF-PSOLPA with direct link and optimal PA at BS and each UAV; and 2) J-HBF-PSOLPA with direct link and optimal PA at each UAV only; 3) J-HBF-PSOLPA without direct link; and 4) J-HBF-FL-EQPA. The analysis can be summarized as follows: 1) the optimal PA from BS and each UAV can provide an improved performance (e.g., an increase of $\approx$ 10\% rate) when compared to only optimal PA at each UAV; 2) by increasing the number of UAVs, we can achieve a higher achievable rate (e.g., the rate can be increased by 5-7$\%$ when $M$ is increased from 2 to 3). Moreover, by increasing the number of UAVs, we can further improve the performance as each UAV cluster less number of users, which leads to reduced MU interference, and thus, an increased achievable rate. However, it must be noted that optimizing the number of UAVs is beyond the scope of this paper, and is left as our future work.
\vspace{-1ex}
\section{Conclusions}
\vspace{-1ex}
In this paper, we considered a MU-mMIMO IoT cooperative relaying system, where multiple UAV DF relays connect the BS to a large number of users. We have proposed a sequential optimization scheme that employed swarm intelligence to assign users to UAVs through K-means clustering, and optimized the UAV locations and power allocation from BS and each UAV, followed by the design of RF and BB stages for maximum achievable rates. The RF stages are designed using the angular information of UAVs and users, while BB stages are designed using reduced-dimension effective channel matrices. Our findings show that multiple UAV-assisted cooperative relaying system works better than single UAV, especially when taking into account the practical user distributions. Moreover, compared to fixed positions and equal power distribution of UAV, the optimization of UAV locations and power allocation substantially improves the achievable rate. 
%\balance
\bibliographystyle{IEEEtran}
\vspace{-2ex}
\bibliography{references}
\vspace{-1ex}
\end{document}